\documentclass[manuscript,screen,nonacm]{acmart}

\usepackage{amsmath}
\usepackage{amsthm}
\usepackage{placeins}

\newtheorem{proposition}{Proposition}

\begin{document}

\title{Can a search hold its own bar? In-loop deflation and its exploits in
autonomous strategy discovery}

\author{Eray Gen\c{c}ay}
\orcid{0000-0002-1510-5628}
\email{eray.gencay@gmail.com}
\affiliation{%
  \institution{Independent Researcher}
  \city{Stuttgart}
  \country{Germany}
}

\begin{abstract}
Corrections for backtest overfitting such as the deflated Sharpe ratio are applied after a search ends: the searcher maximises raw performance and a separate procedure deflates the reported winner. Autonomous searches driven by large language models perform many implicit backtests before reporting, so this paper asks what changes when the correction moves inside the loop and the acceptance threshold is indexed to the running count and dispersion of the search's own trials. The threshold then becomes an object the search can optimise against: we show analytically that a naive in-loop bar admits a variance-dilution effect, in which mediocre near-duplicate candidates lower the bar because the ledger's dispersion falls faster than its trial-count factor grows. We propose two defences, a pre-registered variance floor and a monotone high-water-mark bar, and evaluate the construction under a protocol frozen before any run, with every in-loop decision replaying exactly from its recorded trace. Across 53 audited hundred-candidate searches over two frontier models, two universes, and two feature spaces, the gated search reports ``no certifiable discovery'' on 36 of 37 runs, un-gated controls report a strategy on all 16, and every arm's out-of-sample confidence interval includes zero. Supporting studies find that state-based stopping reaches the same outcome after 8 to 9 trials at about a twelfth of the trial budget while an adaptive budget adds nothing, that a controlled stress test reproduces the dilution effect and both defences, and that a hundred nominal trials collapse to 6 to 8 effectively independent return streams. We scope these results to the analysed system.
\end{abstract}



\maketitle

\section{Introduction}
\label{sec:intro}

A strategy that is chosen as the best of many candidates owes its measured score to two ingredients. One is genuine edge, which tends to repeat on data the search never touched. The other is selection luck, which does not. When a process auditions $N$ candidates and reports the best, the reported score is inflated by the luck that helped put that candidate on top. Empirical finance has long supplied instruments to correct for this, including the deflated Sharpe ratio, the probability of backtest overfitting, and multiplicity-adjusted hurdles \citep{Bailey2014,BaileyLdP2014dsr,Harvey2016}.\footnote{The Sharpe ratio is a strategy's mean return divided by the standard deviation of its returns, annualised throughout this paper. A backtest is a historical simulation of a strategy; point-in-time means universe membership as known on each historical date; SPY is an S\&P~500 index fund, used here as the passive benchmark.} These instruments share a common arrangement. The search runs first, and the correction is applied afterwards, to the reported winner. The searcher itself stays statistically naive: it maximises raw in-sample performance, and honesty is imposed on its output after the fact.

Autonomous search loops built on LLMs make this arrangement uncomfortable. A refinement agent that proposes, backtests, and revises candidate strategies performs its multiplicity inside the loop, quickly, guided by a critic that sees raw performance numbers. In prior work we instrumented such a system and applied the correction after the fact, finding that the apparent edge of every discovered strategy disappeared once its performance was deflated by the search's recorded trial count \citep{Gencay2026}. That verdict is honest, but it arrives only after the search budget has been spent. A natural question follows, and to our knowledge no agent-loop system has made this move: what changes if the correction is moved inside the loop, so that the searcher's own acceptance threshold, stopping rule, and feedback are indexed to the running count and dispersion of its own trials? Optimisation loops for LLM agents are now common, but they do not carry statistical accounting of this kind.

Moving the correction inside the loop is not simply a matter of relocating a computation, and the analytical core of this paper is to explain why. The live threshold, which we call the bar, is $B_k=\sqrt{\widehat V_k}\,K(N_k)$: the dispersion of the trial scores seen so far, multiplied by a factor for the luckiest of $N_k$ draws. Read in words, the bar is an estimate of the best score that selection luck alone would be expected to produce, given how many candidates have been tried and how spread out their scores are. A candidate is accepted only if it clears this bar. Once the bar is part of the searcher's environment, however, it becomes something the search can act upon. The trial-count factor $K(N)$ grows only like $\sqrt{2\ln N}$, whereas a single near-mean trial shrinks the dispersion factor $\sqrt{V}$ at the faster relative rate $1/(2N)$. Adding mediocre, near-duplicate candidates can therefore push the bar down rather than up. We call this variance dilution, state it as a proposition with explicit assumptions, and propose two defences: a pre-registered variance floor that limits how far the bar can be pushed down, and a monotone bar that is not allowed to fall at all.

The empirical program is pre-registered in the sense available to a systems paper. The full protocol, covering parameters, arms, primary endpoints, decision rules, and an acceptable-findings clause, was written into a versioned design document and frozen by commit before any run; the frozen parameters are checked by automated tests; and every in-loop decision of every run can be replayed exactly from its recorded trace. The primary endpoint is deliberately independent of the bar: each arm is judged by an external measure, the out-of-sample Sharpe ratio with a bootstrap confidence interval of whatever that arm selected. This avoids grading a system with the very instrument under study.

We report both what the system rejects and what it certifies. The central result is that the gate changes what is reported rather than what is found. Across 37 gate-active searches (arms whose acceptance objective includes the certification gate), spanning two frontier models, a 453-stock point-in-time equity universe, a 39-ETF transfer universe, and an exploratory fundamentals feature space, 36 end with ``no certifiable discovery,'' while un-gated controls report a selected strategy on every run. In every arm of the study, gated or not, the out-of-sample confidence interval of the selected strategy includes zero. The one certification, at the seed whose in-sample record was strongest, indicates that the gate is discriminating rather than trivially always-empty. Around this central result we place four supporting studies: a comparison of stopping rules, a study of what the critic is allowed to see, a controlled stress test of the dilution effect, and a diagnostic for statistical dependence among the trials. Their findings are summarised in the contributions below and reported in Section~\ref{sec:results}, with claims scoped to the analysed system.

Our contributions are the following.
\begin{enumerate}
\item \textbf{A metric placed inside the search loop, analysed as something the search can game.} We index an LLM search's acceptance threshold to its own running trial ledger, and we show that the naive construction admits a variance-dilution effect. We state the effect as a proposition with listed assumptions (Proposition~\ref{prop:dilution}): to first order, a near-mean trial lowers the bar, because the dispersion factor falls at rate $1/(2N)$ while the count factor recovers only at rate $1/(2N\ln N)$.
\item \textbf{Two defences, with their limits stated.} A pre-registered variance floor bounds the dilution, and a monotone high-water-mark bar removes the incentive to dilute at the cost of some conservatism. We describe the residual incentive that the monotone bar leaves in place rather than claiming it is eliminated.
\item \textbf{A pre-registered, audited demonstration.} Frozen parameters checked by tests; staged execution with a committed audit before each stage; a bar-independent primary endpoint; a bit-exact replay audit of every in-loop decision; and a pre-committed clause under which ``in-loop correction changes nothing measurable'' would be an acceptable result.
\item \textbf{Supporting studies.} State-based stopping obtains the fixed-budget outcome at about one-twelfth of the trial budget on these runs, while adaptive budgeting adds nothing; showing the critic the deflated accounting shifts near-duplication in a direction that flips with the bar mode, consistent with the dilution analysis; the stress test reproduces the effect and both defences; and a dependence diagnostic finds 6 to 8 effectively independent return streams per hundred nominal trials in these searches. A deterministic grid over the frozen hyperparameters re-derives the central findings from the recorded trajectories (Section~\ref{sec:results-sensitivity}).
\end{enumerate}

The remainder of the paper is organised as follows. Section~\ref{sec:related} positions the work. Section~\ref{sec:anatomy} develops the bar and the dilution analysis. Section~\ref{sec:architecture} describes the accounting architecture, and Section~\ref{sec:protocol} the pre-registered protocol. Section~\ref{sec:results} reports results, and Sections~\ref{sec:discussion} to \ref{sec:conclusion} discuss, delimit, and conclude.

\section{Related work}
\label{sec:related}

\paragraph{Backtest overfitting and multiple testing in finance.}
A long line of work quantifies how search inflates backtests. \citet{Bailey2014} define the probability of backtest overfitting and estimate it by combinatorially symmetric cross-validation. \citet{BaileyLdP2014dsr} derive the deflated Sharpe ratio, which tests an observed Sharpe ratio against a threshold that rises with the number of trials and their dispersion. \citet{HarveyLiu2015backtesting} and \citet{Harvey2016} argue that many claimed factors fail appropriately deflated hurdles, and \citet{Arnott2019} and \citet{LopezdePrado2018} codify protocol. The program remains active; \citet{HarveySancettaZhao2026} revisit what threshold should now be applied to factor-model tests given the accumulated mining of the cross-section. Throughout this line, the correction is applied after the search: the search process is untouched, and deflation is applied to its reported winner. Our prior system \citep{Gencay2026} follows the same pattern, instrumenting an LLM search with a complete trial ledger and deflating after the fact. The present paper moves the correction into the search policy itself and studies what that changes, including the new opportunity for gaming that it opens.

\paragraph{Adaptive data analysis and online multiple testing.}
A separate literature asks how statistical validity survives when the analyst adapts to previous results, which is the situation of a search loop. The reusable holdout preserves holdout validity under adaptive querying \citep{Dwork2015}. The Ladder bounds leaderboard error under adaptive submissions by releasing only quantised improvements \citep{BlumHardt2015}, a mechanism our ladder feedback policy adapts for the critic. Online multiple testing supplies a budget formalism: alpha-investing \citep{FosterStine2008}, online false-discovery-rate rules \citep{Javanmard2018}, and their adaptive refinement SAFFRON \citep{Ramdas2018saffron} model an ordered stream of hypotheses that spends and earns significance wealth, which structurally resembles an agent proposing candidates. The field is consolidating around e-values and anytime-valid inference \citep{RamdasWang2025}, with recent work extending online false-discovery-rate control to more general dependence conditions \citep[e-GAI;][]{Zhang2025egai} and to settings where decisions generate feedback that later refines the thresholds \citep{Lu2025gaif}. E-values have entered finance mainly for regulatory validation of risk forecasts, as in e-backtesting of value-at-risk and expected-shortfall forecasts \citep{WangWangZiegel2025}, which shares our concern with optional stopping but addresses forecast calibration rather than search multiplicity. Sequential error control has recently entered agent loops for neighbouring purposes: POPPER runs e-value sequential falsification of an externally supplied hypothesis inside an LLM-agent loop \citep{Huang2025popper}, a functional architecture spends a LORD++ online false-discovery budget inside an AI-scientist loop \citep{Sargsyan2025rigor}, and alpha-investing itself was designed for automated streamwise feature selection, where the threshold adapts inside the selection loop but every test spends fresh data \citep{Zhou2006streamwise}. To our knowledge none of this machinery has been used to correct the selection multiplicity of an autonomous search from inside the loop, where each candidate is scored on the same reused data and the acceptance threshold must answer the search's own running trial count. We borrow its shapes deliberately and state the resulting limits plainly: our alpha-wealth stopping rule is alpha-investing-shaped but, under the dependence present in strategy search, we do not claim a false-discovery-rate guarantee from it (Section~\ref{sec:anatomy-wealth}); the dependence-robust e-value program \citep{Zhang2025egai,RamdasWang2025} is a natural route to such a guarantee and is left to future work.

\paragraph{LLM agents for strategy discovery and alpha mining.}
The system lineage runs from financial LLM infrastructure and memory-augmented trading agents \citep{YangFinGPT2023,YuFinMem2024,ZhangFinAgent2024}, through interactive and self-improving alpha mining \citep{WangAlphaGPT2023,WangQuantAgent2024,YuAlphaGen2023}, to a current generation of full-stack research agents: multi-agent trading desks \citep{Xiao2024tradingagents}, factor and model co-optimisation with bandit-scheduled research loops \citep[RD-Agent(Q);][]{Li2025rdagentq}, and evolutionary trajectory-level search \citep{Han2026quantaalpha}. Closest to our concern, AlphaAgent targets overfitting and alpha decay inside its mining loop through originality enforcement, semantic-rationale checks, and complexity penalties \citep{Tang2025alphaagent}. An agentic factor-investing framework gates candidates on a fixed $t \geq 3.0$ hurdle motivated by the multiple-testing literature \citep{HuangFan2026factor}; the hurdle is set in advance and does not move with the loop's own trial ledger. These are heuristic devices; none indexes the loop's acceptance threshold to the multiplicity of its own search, and reported performance is not deflated by the trial count the loop consumed. Independent evaluations have begun to document the consequences. An audit of execution assumptions across thirty trade-relevant studies finds evaluation reporting lagging architecture reporting, with costs, execution timing, and held-out discipline frequently under-specified \citep{Yao2026beyond}; and a re-evaluation of published LLM timing strategies over two decades and a wider cross-section of symbols finds their reported advantages deteriorate substantially \citep{Li2025finsaber}. Both address the reporting and the sampling sides of the problem, and neither corrects for the multiplicity of the search that produced the result. That gap is the subject of this paper. It is worth noting that each new system in this line tends to be more autonomous and productive than its predecessor, which tends to enlarge the implicit trial count behind reported backtests.

\paragraph{Agents that game their evaluations.}
A growing literature on evaluation integrity documents that LLM agents can exploit the tests by which they are scored. Given access to unit tests, models may delete or special-case failing assertions rather than fix the underlying code \citep{Zhong2025impossiblebench}, systematic red-teaming audits find exploitable flaws across major agent benchmarks and argue that evaluations should be secure by design \citep{Wang2026benchjack}, and purpose-built benchmarks measure how readily tool-using agents exploit such shortcuts \citep{Thaman2026rhb}. Our variance-dilution analysis (Section~\ref{sec:anatomy}) is a statistical counterpart to this phenomenon, derived rather than observed. Once the correction is part of the searcher's environment, the bar is something the search can optimise against, and a budgeted stopping rule can, in principle, help fund the exploit. In this vocabulary, the hardened bars we pre-register are a form of defence-aware evaluation design, and our feedback and stress-test results (Sections~\ref{sec:results-e3} and \ref{sec:results-e4}) measure both how readily the exploit is learned and how effectively the defences close it.

\paragraph{Position.}
The neighbouring literatures each provide one ingredient: post-hoc deflation in finance, adaptive-validity mechanisms in statistics, autonomous strategy searchers among LLM agents, and evidence of evaluation gaming in agent evaluation. We are not aware of prior work that places the multiplicity correction inside an autonomous search loop, analyses the loop's incentive to game it, defends against that incentive, and measures the result under a pre-registered protocol. That combination is the contribution.

\section{The bar and how a search can game it}
\label{sec:anatomy}

\subsection{Ledger and bar}
\label{sec:anatomy-bar}

The searcher maintains a trial ledger. After $k$ refinement steps it has performed $N_k$ design-window evaluations, comprising the initial plan, every candidate that backtested, and any pre-existing seed trials, with annualised design Sharpe ratios whose sample variance is $\widehat V_k$. A candidate's own evaluation is counted in the ledger at the moment it is judged. The design window is the in-sample period the search optimises on; the evaluation window is the held-out period, used only by the external endpoint of Section~\ref{sec:protocol}. The ledger opens with a small baseline, the initial plan and the pre-existing seed trials, so a run that spends its full budget of 100 refinement candidates records about a hundred trials, typically 102 to 104 on these runs, and counts quoted in refinement candidates exclude that baseline. Baseline trials without a recorded design Sharpe ratio enter the count $N_k$ but not the dispersion $\widehat V_k$, a convention that can only raise the bar on these runs. The live bar is the expected-maximum threshold of \citet{BaileyLdP2014dsr}, evaluated at the current ledger state:
\begin{equation}
B_k \;=\; \sqrt{\max(\widehat V_k,\, V_{\mathrm{floor}})}\; K(N_k),
\qquad
K(N) = (1{-}\gamma)\,\Phi^{-1}\!\Big(1{-}\tfrac1N\Big) + \gamma\,\Phi^{-1}\!\Big(1{-}\tfrac{1}{Ne}\Big),
\label{eq:bar}
\end{equation}
where $\Phi$ is the standard normal distribution function and $\gamma$ the Euler--Mascheroni constant. In words, $B_k$ estimates the best Sharpe ratio that selection luck alone would be expected to produce, given how many candidates have been tried and how dispersed their scores are: $\sqrt{V}$ sets the scale, and $K(N)$ counts how many scale units above the mean the luckiest of $N$ noise draws typically lands. A candidate with raw metric $s$ is certified when $s > B_k$. The variance floor $V_{\mathrm{floor}}$ is a pre-registered constant whose role is explained in Section~\ref{sec:anatomy-dilution}; setting it to zero recovers the naive bar, which we call the plain bar.

\subsection{Certifiable improvement, stopping, and the honest empty}
\label{sec:anatomy-cert}

A threshold common to all candidates does not change which candidate ranks best, only whether the best one is accepted. The loop's selection rule is therefore certifiable improvement: a candidate becomes the incumbent only if it beats the running best and also clears the bar at its own evaluation. Because the bar moves as the ledger grows, certification at evaluation time is provisional. At termination the reported candidate is re-checked against the final ledger, and a run whose best candidate fails the final bar reports ``no certifiable discovery.'' We treat this as a first-class outcome and call it an honest empty. Stopping is the legitimate way to freeze the bar: a search that stops early locks in a lower threshold, which is the appropriate behaviour for a certifier that has decided to stop looking. We therefore treat the stopping rule as part of the policy under study. Three rules are considered: a fixed budget (the control); an evaporation rule that stops once the bar has stayed at or above the running best for $m$ consecutive evaluations, that is, once a further trial would on balance cost more evidence than it can supply; and an alpha-wealth rule described in Section~\ref{sec:anatomy-wealth}.

\subsection{Variance dilution}
\label{sec:anatomy-dilution}

Once the bar is part of the searcher's environment, its two factors respond in opposite directions to an unremarkable trial. The following proposition makes this precise for the idealised case of a trial placed exactly at the ledger mean.

\begin{proposition}[Variance dilution]
\label{prop:dilution}
Assume that (i) the ledger already contains $N$ trials with sample variance $V>0$; (ii) one further trial is appended whose score equals the current ledger mean; and (iii) the count factor is well approximated in the relevant range by $K(N)\approx\sqrt{2\ln N}$, treated as differentiable in $N$. Then, expanding each factor to first order in $1/N$, appending the trial changes the logarithm of the bar $B=\sqrt{V}\,K(N)$ by
\begin{equation}
\Delta \log B \;\approx\; -\frac{1}{2N}\Big(1 - \frac{1}{\ln N}\Big),
\label{eq:dilution}
\end{equation}
which is negative for $N>e$. Consequently, for $N>e$, a mean-valued trial lowers the bar.
\end{proposition}

\begin{proof}[Proof sketch]
Appending a trial at the mean leaves the sum of squared deviations unchanged and increases the denominator, so the sample variance scales by $(N-1)/N$ and the dispersion factor changes relatively by $\sqrt{(N-1)/N} = (1-1/N)^{1/2} \approx 1 - 1/(2N)$. For the count factor, differentiating $K(N)=\sqrt{2\ln N}$ gives $K'(N)/K(N) = 1/(2N\ln N)$. Summing the two relative changes yields \eqref{eq:dilution}. The bracket is positive for $\ln N > 1$, i.e.\ $N>e$.
\end{proof}

The two rates differ only by the factor $\ln N$ in the denominator of the count term, so the dispersion effect dominates. The practical consequence, which we measure directly in Section~\ref{sec:results-e4}, is that beyond a few dozen trials a candidate within roughly one ledger standard deviation of the mean tends to lower the bar, while only tail candidates raise it. Mediocre candidates are thus not neutral; they can help certify the incumbent. A search that discovers this can lower its own bar by variance dilution, that is, by padding the ledger with near-mean, near-duplicate candidates so that $\sqrt{\widehat V}$ falls faster than $K(N)$ rises. In this sense a naive in-loop bar is gameable.

Two remarks bound the claim. The proposition is a first-order statement under the stated approximations, chiefly the near-mean placement (ii) and the asymptotic form of $K$ (iii); the exact two-term $K(N)$ in \eqref{eq:bar} shifts the constants but not the sign for the range of $N$ we study. And the result describes an incentive gradient, not an inevitability: whether a given searcher acts on it is an empirical question, which the feedback study (Section~\ref{sec:results-e3}) addresses.

Two defences enter the specification, together with a filter. First, a variance floor $V_k=\max(\widehat V_k, V_{\mathrm{floor}})$ bounds how far dilution can lower the bar and also stabilises the bar at small $k$; $V_{\mathrm{floor}}$ is calibrated once, from prior-system search ledgers only (Section~\ref{sec:protocol}). Second, a high-water-mark bar $B^{\mathrm{hwm}}_k=\max_{j\le k}B_j$ is not allowed to fall, which removes the downward movement that the exploit relies on, at the cost of some conservatism. This defence is not absolute: while holding an incumbent, a search still has a mild incentive to add bar-neutral filler rather than exploratory candidates, since under a monotone bar exploration can only push the permanent bar up and threaten the final re-check. That residual incentive is bounded, because filler pays the per-trial cost and can never produce the new record that certification requires, and it is observable through the duplicate-rate endpoint. The filter, a precondition rather than a defence: exact duplicates are rejected before evaluation, which removes the zero-information limit of the attack; near-duplicates are genuine, if correlated, trials and are evaluated.

\subsection{An adaptive budget, and how dilution could fund itself}
\label{sec:anatomy-wealth}

The alpha-wealth stopping rule is alpha-investing-shaped \citep{FosterStine2008}: it begins with wealth $W_0$, charges a toll $\varphi$ per evaluation, pays a reward $\psi$ for each certified-at-evaluation improvement, and stops when wealth reaches zero. We do not claim a false-discovery-rate guarantee from it. Under the dependence present in strategy search the guarantees of online false-discovery-rate theory do not transfer, so we treat wealth simply as an adaptive budget in which success lengthens the search and failure shortens it, and we ask experimentally whether that adaptivity is worth anything relative to state-based stopping. Rewards are earned at evaluation time and are not clawed back by the final re-check, since a search cannot spend, at step $k$, wealth whose validity depends on trials after $k$. This budget inherits the dilution incentive by composition: under a plain bar, dilution lowers the bar, easier certification earns $\psi$, and $\psi$ buys further iterations in which to dilute. In principle the exploit could therefore help fund itself. Whether it does so in practice depends on the budget parameters, and we return to this in Section~\ref{sec:results-e4}, where at our frozen parameters it does not. A monotone bar removes the downward movement and, with it, this funding channel.

\subsection{What the critic is allowed to see}
\label{sec:anatomy-feedback}

The third policy axis is the information shown to the proposing critic. Full feedback shows raw design metrics, as in the prior system. Deflated feedback additionally shows the live accounting: the bar, the excess over it, trials spent, and wealth remaining. Ladder feedback shows only quantised information in the style of \citet{BlumHardt2015}, namely whether the candidate improved on the running best by at least a step $\eta$, which bounds how much the search can learn from its own evaluations. Deflated feedback makes the loop statistically transparent, and by the analysis above transparency can act in either direction. A critic that sees the bar might steer away from duplicates, because they cost budget and cannot certify until the bar has fallen; or it might infer the dilution gradient and produce them. Which effect dominates is an empirical question, and the analysis predicts that its sign should depend on the bar mode, since a monotone bar offers no downward gradient to exploit. We pre-register this as a two-sided prediction and test it in Section~\ref{sec:results-e3}.

\subsection{Dependence and effective ledger size}
\label{sec:anatomy-neff}

Correlated trials affect the bar in two opposing ways. They reduce the number of effectively independent draws, so that $K(N)$ evaluated at the nominal $N$ overstates the luck ceiling, which is conservative; but clusters of similar trials also deflate the measured dispersion $\widehat V$, which is the dilution mechanism itself. A natural idea is to run the bar at an effective count such as the Kish size $N_{\mathrm{eff}}=N/(1+(N-1)\bar\rho)$ \citep{Kish1965}, where $\bar\rho$ is the mean pairwise correlation. We do not do this in the loop, because it would make the bar easier to game rather than harder: a near-duplicate flood raises $\bar\rho$, which lowers $N_{\mathrm{eff}}$ and hence $K(N_{\mathrm{eff}})$, while the same clones lower $\widehat V$, so both factors of the bar would fall together. Using the nominal $N$ is, by contrast, a partial defence, since its $K(N)$ rises under the flood and opposes the dispersion effect. The live bar therefore always uses nominal $N$. We compute effective size (Kish size and an eigenvalue-based effective rank) only once, after the fact and identically for every arm, from the final candidate-return matrix, and we report the resulting gap between the nominal and effective bars as a diagnostic that no arm can influence in its own favour (Section~\ref{sec:results-neff}). A properly in-loop treatment of dependence is better addressed through the e-value program \citep{RamdasWang2025,Zhang2025egai}, which we leave to future work.

\section{Accounting architecture}
\label{sec:architecture}

The loop under study is the refinement loop of the prior system \citep{Gencay2026}, inherited unchanged apart from the accounting surface: a proposer writes a candidate, a declarative trading-strategy plan over registered features; the plan is backtested on the design window and its annualised Sharpe ratio enters the trial ledger; a critic sees the configured feedback block and steers the next proposal. The accounting surface sits between those steps: the objective computes value, bar, and certification from the ledger; the feedback policy renders what the critic may see; the stopping rule decides whether the search continues; and the external endpoint of Section~\ref{sec:protocol} later judges whatever the run reported.

The constraint that shapes the implementation is auditability: every in-loop decision must be reconstructible, exactly, from the run's recorded trace. Three properties provide this.

\paragraph{Registries of pure policies.}
Objectives, stopping rules, and feedback policies are registry entries, so that adding one is a matter of adding a definition rather than editing the loop. Each entry is a pure function of the recorded ledger state. The objective maps candidate metrics, ledger state, and parameters to a verdict of value, threshold, excess, and certification; the stopping rule maps ledger state and parameters to a decision with a recorded reason; the feedback policy renders the critic-visible block from the same state. No entry holds hidden state. Parameters are validated when the configuration is parsed, so that unknown keys are rejected and policies that require a bar (evaporation, wealth certification) are rejected when paired with a bar-free objective. A misconfigured protocol thus fails before any run begins.

\paragraph{Back-compatibility invariant.}
The default configuration, which uses the raw metric, a fixed budget, and full feedback, reproduces the prior system's loop exactly and emits a byte-identical trace on a pinned reference run. This is load-bearing rather than cosmetic, because the control arm of every experiment is the prior system, so the invariant is what lets us attribute a treatment effect to the accounting surface and not to an incidental change. The instrumented surface activates only when a configuration departs from the default along a declared axis.

\paragraph{Bit-exact replay audit.}
A replay auditor reconstructs the entire accounting, including every bar value, excess, certification, wealth balance, and stopping decision, from the trace's recorded candidate metrics alone, and compares it field by field with what the live loop recorded. We require exact equality rather than tolerance, since the replay repeats the same floating-point operations in the same order on the same inputs. Every audited run in this paper, including every live LLM run, replays cleanly, and a doctored trace is detectable by construction. Replay verifies the accounting conditional on the recorded proposals; the proposals themselves are sampled at default temperature and ship as recorded traces, not as something regenerable. Provenance is enforced operationally as well: every suite run carries a machine-readable tag, and a guard aborts any suite-tagged run whose configured model differs from the frozen set before any spend occurs.

\section{Pre-registered protocol}
\label{sec:protocol}

\subsection{Freeze discipline}
The full protocol, covering parameters, arms, endpoints, decision rules, staging, and the falsification clause, was written into a versioned design document and frozen by commit before any suite run. That commit is the internal pre-registration boundary. The few post-freeze amendments are dated and justified in place: a feedback-delivery clarification, the exploratory fundamentals cell, and implementation notes that fixed measurement choices before any result was inspected. Two mechanisms make the freeze operational rather than nominal. First, the frozen cells are checked by automated tests: the committed test suite asserts the exact set of cells, pins each cell's full accounting policy so that every cell differs from its reference arm along exactly its declared axis, asserts that each treatment configuration departs from the control only inside the accounting block, and reads frozen constants such as the variance floor from the calibration record rather than re-typing them. Second, execution was staged, with each stage's outputs audited for grid completeness, frozen-model provenance, control back-compatibility, and bit-exact replay of every accounting decision, and with the audit report committed before the next stage began.

\subsection{Calibration and frozen parameters}
The variance floor is fixed by prior data and not by any outcome of this study. We set $V_{\mathrm{floor}}=0.120$ in annualised Sharpe variance, the tenth percentile of the running trial-Sharpe variances $V_k$ for $k\ge10$, pooled across the eight pre-registered search ledgers of the prior system. Table~\ref{tab:params} lists the frozen parameters, and Table~\ref{tab:design} summarises the experimental cells. A cell is an arm crossed with a model. For E2 and E3 the reference configuration is the gate arm of E1 rather than the un-gated control, so each cell still differs from its reference along one declared axis; the gate-active count is therefore $10+6+15+3+3=37$ across E1, E2, E3, the transfer gate, and the exploratory gate. The exploratory fundamentals pair probes the gate where the control's selections look strongest out of sample; it is a post-freeze amendment, reported separately throughout, and included in the run-level counts.

\begin{table}[htbp]
\centering
\caption{Frozen protocol parameters, fixed before any suite run.}
\label{tab:params}
\footnotesize
\begin{tabular}{ll}
\toprule
Search budget & $100$ refinement candidates per run \\
Models & GPT-4.1 (all cells); Claude Sonnet 5 (headline cells) \\
Seeds & 3 (comparisons); 5 (robustness, headline pair) \\
Variance floor $V_{\mathrm{floor}}$ & $0.120$ (ann.\ Sharpe var.; 10th pct.\ of prior-system $V_k$) \\
Evaporation patience $m$ & 3 (fixed per bar mode) \\
Ladder step $\eta$ & 0.10 \\
Wealth $(W_0,\varphi,\psi)$ & $(100, 1, 20)$; a zero-certification run spends $100$ evaluations \\
Universe and windows & 453 US equities and SPY; design 2017--2021, evaluation 2022--2025 \\
Transfer universe & 39 ETFs; design 2007--2016, evaluation 2017--2025 \\
\bottomrule
\end{tabular}
\end{table}

\begin{table}[htbp]
\centering
\caption{Experimental cells. Each cell equals its reference arm except along its declared axis (the un-gated control for E1 and E5; the E1 gate arm for E2 and E3), which is checked by committed tests. Runs count seeds per arm, summed over models where both run ($5{+}5$: five seeds on each of two models; $4{\times}3{+}3$: four GPT-4.1 arms at three seeds plus one Claude cell at three). E4 is deterministic and uses no LLM. The fundamentals pair is exploratory, reported separately from the frozen cells.}
\label{tab:design}
\footnotesize
\begin{tabular}{llll}
\toprule
Experiment & Perturbed axis & Arms & Runs \\
\midrule
C (control) & none (prior system) & raw metric, fixed 100, full feedback & $5{+}5$ \\
E1 (gate) & objective & deflated certification gate & $5{+}5$ \\
E2 (stopping) & stopping rule & evaporation$(3)$; alpha-wealth$(100,1,20)$ & $3{+}3$ \\
E3 (feedback, bar) & critic surface, bar mode & full/deflated/ladder, plain; deflated, hwm & $4{\times}3{+}3$ \\
E4 (dilution stress) & scripted near-mean flood & plain/floored/hwm, evap./wealth & $12$ streams \\
E5 (robustness) & universe transfer & 39-ETF control and gate & $3{+}3$ \\
Exploratory & feature space & fundamentals control and gate & $3{+}3$ \\
\bottomrule
\end{tabular}
\end{table}

\subsection{Endpoints and falsification}
The primary endpoint is independent of the bar and identical for every arm: for each run, the annualised out-of-sample Sharpe ratio of the selected strategy, with a stationary-bootstrap 95\% confidence interval on the held-out window. The bootstrap is stationary with mean block length near $n^{1/3}$, 1000 resamples, percentile intervals, and a fixed seed \citep{PolitisRomano1994}. For a gate run that certifies nothing, the endpoint evaluates the strategy the run would otherwise have reported, the last candidate certified at its own evaluation or the initial plan when there was none; the honest-empty label applies to the certification verdict, not to the existence of a selection. The endpoint answers a single question, whether the thing the search selected survived on data the search never used. Secondary bar-free endpoints are the in-sample-to-out-of-sample retention, trials consumed, near-duplicate rates, distinct-hypothesis count, the rate of ``no certifiable discovery,'' and the gap between the nominal and effective bars. Arm-internal quantities, including certification, the deflated Sharpe ratio, the probability of backtest overfitting, wealth trajectories, and stop reasons, are reported as descriptive detail, and no claim rests on any arm's own bar. We committed in advance to a falsification clause: if the gate neither improves out-of-sample calibration nor produces honest empties, and the feedback axis shows no behavioural difference, and the dilution analysis fails to reproduce, then in-loop correction would be inert and we would report exactly that. Reporting is symmetric by construction, in that every arm appears in every endpoint table and ``no certifiable discovery'' is a reported row rather than a blank.

\subsection{Execution}
All 53 audited runs completed, with every accounting decision replaying exactly, in four stages with committed audits between them. Two labelled pilot runs preceded the suite for cost calibration and enter no table. Total application-programming-interface spend was approximately \$160 by provider billing, split as about \$69 for the OpenAI model and about \$88 for the Anthropic model. The accounting arms' and pilots' spend is also recorded per run in the shipped manifests, about \$102 in total; the un-gated controls run the prior system's uninstrumented path and record no usage, and the committed cost record reconciles the recorded and billed figures at the measured per-run rates. A single hundred-candidate search cost about \$1.8 on GPT-4.1 or about \$6.8 on Claude Sonnet 5, and took roughly 28 to 80 minutes of wall-clock time by the operational logs.

\FloatBarrier
\section{Results}
\label{sec:results}

Figures and tables are generated by a single deterministic script from the committed endpoint records, so that every reported number regenerates from source. We report medians across seeds with seed ranges, and we frame the small-sample comparisons as directions rather than significance claims.

\subsection{The gate changes what is reported, not what is found}
\label{sec:results-e1}

\begin{figure*}[htbp]
\centering
\includegraphics[width=\textwidth]{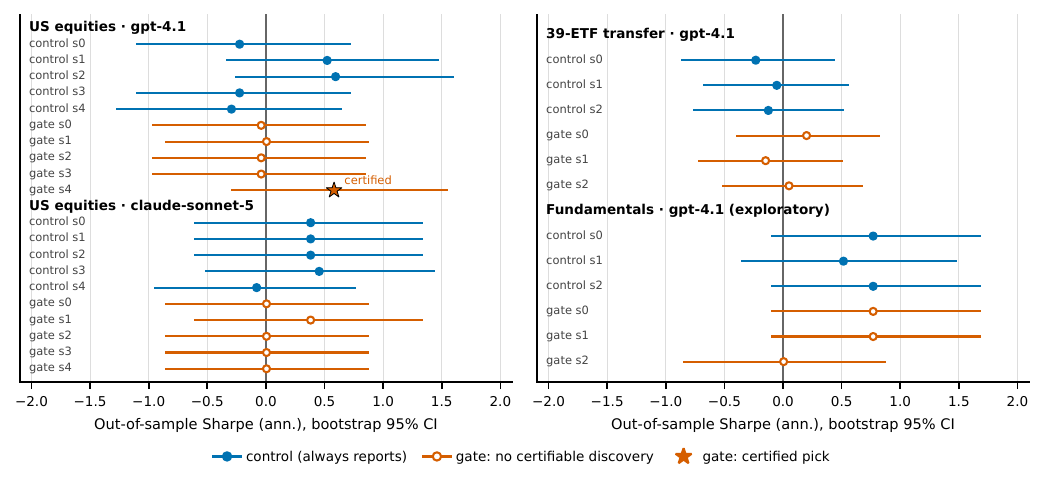}
\caption{Out-of-sample record of every selected strategy, by context and arm.
Controls (filled) report a strategy on every run; the gated search (open
circles) reports ``no certifiable discovery'' on 15 of the 16 gate runs shown,
and the star marks the single certified strategy. Every confidence interval
includes zero. The fundamentals block is exploratory and is reported separately from
the frozen cells.}
\Description{Forest plot of out-of-sample Sharpe ratios with bootstrap
confidence intervals for every control and gate run, in four panels by
universe and model. Every interval crosses zero. Filled circles mark control
picks, open circles mark gate runs reporting no certifiable discovery, and a
star marks the single certified pick.}
\label{fig:forest}
\end{figure*}

Figure~\ref{fig:forest} shows the primary endpoint for every control and gate run across the four contexts, and Table~\ref{tab:endpoints} gives the full endpoint set per cell. The pattern is consistent. The un-gated control reports a selected strategy on all 16 of its runs, and those strategies are noise-shaped: the median out-of-sample Sharpe ratio is $-0.22$ on US equities under GPT-4.1, with a seed range of $-0.29$ to $+0.59$; $+0.38$ under Claude Sonnet 5; and $-0.13$ on the ETF transfer universe. In every arm of the study the 95\% confidence interval of the selected strategy includes zero, so the share of runs whose interval excludes zero is $0.00$ in every row of Table~\ref{tab:endpoints}. Facing the same data, per-seed initial plans, and search budget, the gated search reports ``no certifiable discovery'' on 36 of its 37 runs.

The single exception, on US equities under GPT-4.1 at the seed whose in-sample record was strongest, is certified, and its out-of-sample Sharpe ratio is $+0.58$ with an interval of $[-0.29, 1.54]$. This one certification is informative: a gate that never certified anything could not be distinguished from a constant, whereas one certification out of 37, at the strongest seed, indicates that the gate discriminates while almost always declining. Its interval, like every other in the study, includes zero; certification asserts that the in-sample record cleared the multiplicity-adjusted bar, not that the held-out window can confirm the edge. Certification is thus a claim about the in-sample record relative to the search's own multiplicity; on these runs the one certified record was not confirmed out of sample, which is what symmetric reporting should show when nothing confirmable is present. At any fixed ledger state the bar does not re-rank candidates (Section~\ref{sec:anatomy-cert}); because the bar moves as the ledger grows, acceptance is decided per candidate at its own evaluation, and a gate run reports its last certified incumbent or its initial plan. What the gate changed is the system's willingness to vouch for what its search produced, and out of sample that reticence was not contradicted by any interval. The pattern holds on both models and on the ETF transfer universe. In the exploratory fundamentals space, reported separately, the control selects strategies with out-of-sample Sharpe ratios up to $+0.77$ that the gate declines to certify on all three runs; we treat this as suggestive rather than confirmatory. The strongest frozen-cell median, $+0.76$ in the full-feedback plain-bar cell, carries the same reading: every seed's out-of-sample interval includes zero, so the gate's refusals are not contradicted there either.

\begin{table*}[tp]
\centering
\caption{Bar-free endpoints per cell (median across seeds; brackets give the seed range for the primary endpoint). CI$\not\ni$0 is the share of runs whose out-of-sample interval excludes zero. Cert.\ is the share of gate-active runs certified against the final ledger (not applicable for un-gated controls). Dup(corr) is the near-duplicate rate at nearest-prior return correlation at least $0.95$. Kish $N_{\mathrm{eff}}$ is the effective trial count of the run's ledger; the ranges quoted in the text cover the roughly hundred-trial runs. Exploratory rows are shown separately.}
\label{tab:endpoints}
\footnotesize
\setlength{\tabcolsep}{4pt}
\resizebox{\textwidth}{!}{\begin{tabular}{llrlrrrrrr}
\toprule
Cell & Model & n & OOS Sharpe med [min, max] & CI$\not\ni$0 & Certified & Trials & Dup(corr) & Distinct hyp. & Kish $N_{\mathrm{eff}}$ \\
\midrule
c\_control & claude-sonnet-5 & 5 & +0.38 [-0.08, +0.45] & 0.00 & n/a & 99 & 0.43 & 44 & 22.6 \\
c\_control & gpt-4.1 & 5 & -0.22 [-0.29, +0.59] & 0.00 & n/a & 103 & 0.62 & 25 & 23.4 \\
e1\_gate & claude-sonnet-5 & 5 & +0.01 [+0.01, +0.38] & 0.00 & 0.00 & 99 & 0.42 & 44 & 20.0 \\
e1\_gate & gpt-4.1 & 5 & -0.04 [-0.04, +0.58] & 0.00 & 0.20 & 102 & 0.63 & 26 & 24.6 \\
e2\_alpha\_wealth & gpt-4.1 & 3 & +0.10 [-0.04, +0.10] & 0.00 & 0.00 & 103 & 0.63 & 26 & 32.2 \\
e2\_evaporation & gpt-4.1 & 3 & +0.10 [-0.44, +0.10] & 0.00 & 0.00 & 8 & 0.00 & 6 & 6.0 \\
e3\_deflated\_hwm & claude-sonnet-5 & 3 & +0.01 [+0.01, +0.01] & 0.00 & 0.00 & 98 & 0.44 & 40 & 19.4 \\
e3\_deflated\_hwm & gpt-4.1 & 3 & +0.10 [+0.01, +0.70] & 0.00 & 0.00 & 103 & 0.64 & 23 & 17.4 \\
e3\_deflated\_plain & gpt-4.1 & 3 & +0.10 [+0.01, +0.59] & 0.00 & 0.00 & 103 & 0.69 & 24 & 16.5 \\
e3\_full\_plain & gpt-4.1 & 3 & +0.76 [+0.59, +0.77] & 0.00 & 0.00 & 103 & 0.65 & 26 & 19.1 \\
e3\_ladder\_plain & gpt-4.1 & 3 & +0.01 [+0.01, +0.22] & 0.00 & 0.00 & 103 & 0.64 & 22 & 14.2 \\
e5\_etf\_control & gpt-4.1 & 3 & -0.13 [-0.23, -0.05] & 0.00 & n/a & 104 & 0.66 & 25 & 11.6 \\
e5\_etf\_gate & gpt-4.1 & 3 & +0.05 [-0.15, +0.20] & 0.00 & 0.00 & 104 & 0.61 & 26 & 10.2 \\
\midrule
\multicolumn{10}{l}{\emph{Exploratory (fundamentals)}} \\
\midrule
x\_fund\_control & gpt-4.1 & 3 & +0.77 [+0.52, +0.77] & 0.00 & n/a & 93 & 0.65 & 25 & 11.0 \\
x\_fund\_gate & gpt-4.1 & 3 & +0.77 [+0.01, +0.77] & 0.00 & 0.00 & 96 & 0.64 & 26 & 11.9 \\
\bottomrule
\end{tabular}
}
\end{table*}

\FloatBarrier
\subsection{Honest stopping is inexpensive on these runs}
\label{sec:results-e2}

\begin{figure}[htbp]
\centering
\includegraphics[width=\columnwidth]{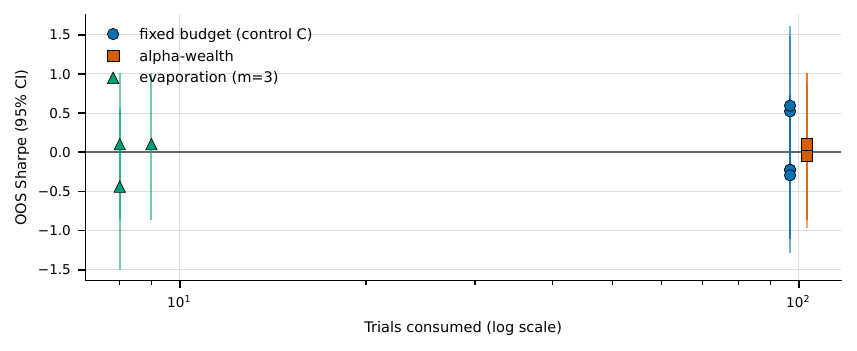}
\caption{Trials consumed against out-of-sample outcome by stopping rule
(GPT-4.1; vertical bars are 95\% confidence intervals; fixed-budget points are
offset horizontally for visibility). Evaporation stops at 8 to 9 trials with
outcomes whose intervals overlap those of the 103-trial arms; alpha-wealth reaches
the 100-iteration cap and consumes the full fixed budget.}
\Description{Scatter plot of trials consumed on a logarithmic axis against
out-of-sample Sharpe ratio with confidence intervals, for three stopping
rules. Evaporation points sit at 8 to 9 trials; fixed-budget and alpha-wealth
points sit at about 100 trials. The confidence intervals of all three rules
overlap.}
\label{fig:stopping}
\end{figure}

Figure~\ref{fig:stopping} plots trials consumed against the out-of-sample outcome for the three stopping rules on GPT-4.1. The evaporation rule stops after 8 to 9 trials on every seed, because the bar, fed by the search's own early trials, rises above the running best within the first handful of evaluations and stays there. It consumes about one-twelfth of the fixed budget, with recorded interface spend of roughly \$0.10 against \$1.8 for a full run, and selects strategies whose out-of-sample intervals overlap those of the fixed-budget arm. The alpha-wealth budget did not stop early. Each of the three seeds earned exactly one reward for a candidate that was certified at its own evaluation but revoked by the final re-check, so wealth stood at 20 rather than 0 when the loop reached its 100-iteration cap, and all three consumed the full fixed budget of 103 trials. On these runs, then, adaptivity of the budget did not trigger any early stop, whereas making the stopping rule read the bar rather than a wallet captured most of the available saving. For a search whose honest verdict is that nothing certifies, the same instrument that declines to vouch also indicates, early and legibly, that continuing is not worthwhile. We state this as a property of these runs rather than as a general result.

\FloatBarrier
\subsection{What the critic sees shifts near-duplication, and the shift depends on the bar}
\label{sec:results-e3}

\begin{figure}[htbp]
\centering
\includegraphics[width=\columnwidth]{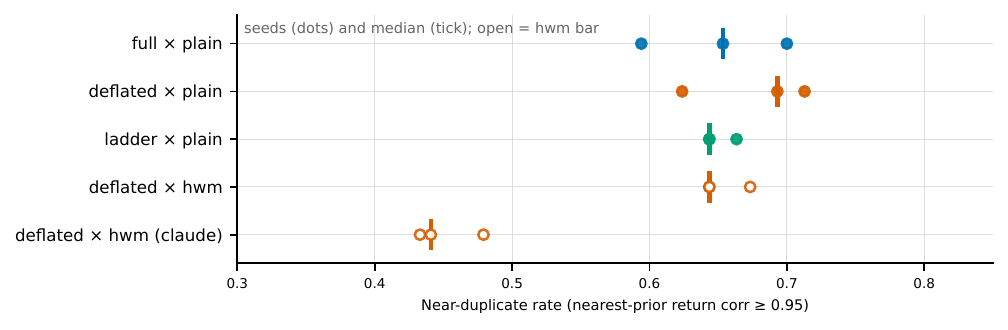}
\caption{Near-duplicate rate by feedback policy and bar mode (dots are seeds,
the tick is the median, open markers are the monotone bar). Deflated feedback
is associated with more duplication under the plain bar and with baseline
duplication under the monotone bar, the direction predicted in
Section~\ref{sec:anatomy-feedback}.}
\Description{Dot plot of near-duplicate rates for five feedback and bar-mode
combinations, with per-seed dots and a median tick per row. Deflated feedback
under the plain bar shows the highest median rate, deflated feedback under
the monotone bar returns to the baseline, and the Claude row sits well below
every GPT-4.1 row.}
\label{fig:dup}
\end{figure}

Figure~\ref{fig:dup} shows the near-duplicate rate, defined as the fraction of evaluated candidates whose design-window returns correlate at least $0.95$ with an earlier trial, across the feedback and bar-mode grid. The pre-registered prediction (Section~\ref{sec:anatomy-feedback}) was two-sided, with a sign that should depend on the bar mode, and the data are consistent with it. Under the plain bar, which can fall, showing the critic its own deflated accounting is associated with a higher median duplicate rate than full feedback, $0.69$ against $0.65$, the direction in which a transparent loop would drift toward the behaviour the dilution analysis rewards. Under the monotone bar, which cannot fall, the same deflated feedback returns to the baseline rate, $0.64$, consistent with the downward gradient having been removed. Ladder feedback sits at the baseline, $0.64$, with the least information released. With three seeds per cell we report these as distributions and directions and do not attach significance to the differences; the direction pattern, however, matches the pre-registered prediction, and it carries the design implication discussed in Section~\ref{sec:discussion}, that transparency about the bar is safe only when the bar cannot be pushed down. We also note a level difference between models: the Claude searches duplicate markedly less on the same cell, $0.44$, and explore roughly twice as many distinct hypotheses per hundred trials (Table~\ref{tab:endpoints}), which indicates that duplication pressure reflects the searcher as well as the incentive. The feedback grid itself runs on GPT-4.1, with Claude replicating one cell, so the sign pattern is a one-model observation.

\FloatBarrier
\subsection{The dilution effect, reproduced and contained}
\label{sec:results-e4}

\begin{figure}[htbp]
\centering
\includegraphics[width=\columnwidth]{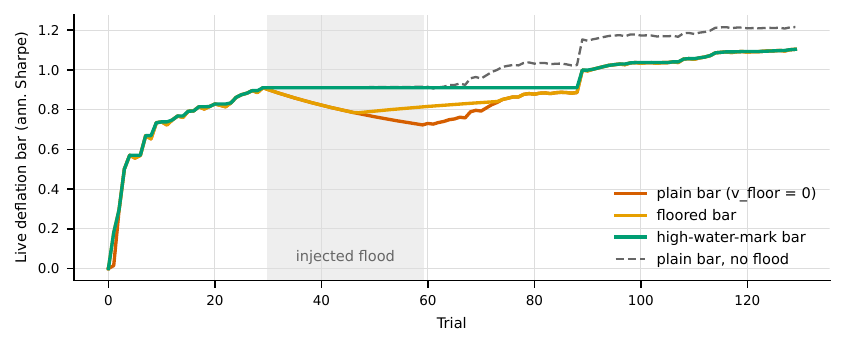}
\caption{One representative real trial stream under the scripted near-mean
flood (shaded), with alpha-wealth stopping. The plain bar falls, the variance
floor arrests the fall, and the monotone bar does not move down; the dashed
no-flood reference is shifted horizontally so that identical real trials align.}
\Description{Line chart of the live deflation bar over about 130 trials for
four bar variants. Inside the shaded flood window the plain bar falls, the
floored bar falls less and recovers once the floor binds, the high-water-mark
bar never moves down, and the dashed no-flood reference continues to rise.}
\label{fig:flood}
\end{figure}

The stress test is deterministic. Recorded real trial streams, comprising the two long prior-system ledgers and this study's ten control ledgers, are replayed through the accounting with a scripted flood of thirty near-mean, never-exact duplicates injected at trial 30, under all three bar modes and both adaptive stopping rules. The driver's incumbent rule is at least as strict as the live loop's, so its self-funding measurements read as lower bounds. Figure~\ref{fig:flood} shows one representative stream, and Table~\ref{tab:e4} sets the pre-registered claims against the measurements across all twelve streams. The mechanical predictions of Section~\ref{sec:anatomy-dilution} are reproduced: the plain bar falls under the flood in every flooded cell, by a median of $0.18$ in annualised Sharpe over the flood window; the variance floor arrests the fall, to a median of $0.05$, with the bar often rising once the floor binds because only $K(N)$ still moves; the monotone bar does not fall in any cell; and the flood never earns wealth, since a near-mean candidate can never be the new record that certification requires.

Two further findings complete the mechanism. First, early stopping removes the opportunity for the attack: in every evaporation cell the search stops before trial 30, because on real streams the bar overtakes the running best well before the injection point, so the flood never executes. An attack on the bar presupposes a search that is still running, and state-based stopping tends to remove that precondition. Second, at our frozen budget the exploit does not fund itself: the lowered plain bar does produce some extra post-flood certifications, the channel that Section~\ref{sec:anatomy-wealth} describes; Table~\ref{tab:e4} shows four post-flood incumbents across the twelve streams under the plain bar against one each under the floored and monotone bars. But the reward earned does not repay the thirty tolls the flood costs, so on these streams dilution never buys runway, and on the streams where it changes the trial count at all it costs 10 to 27 trials. We therefore report the effect as mechanically real and, under these defences and this budget, contained; we do not claim it is harmless in general, and a different budget could tip the balance.

\begin{table}[htbp]
\centering
\caption{Dilution stress test: pre-registered claims against measurement over
twelve real trial streams. Evaporation is assessed per stream and bar mode,
36 cells in all. The certification and runway rows record the self-funding
channel of Section~\ref{sec:anatomy-wealth}, which does not open at the
frozen budget.}
\label{tab:e4}
\footnotesize
\begin{tabular}{lr}
\toprule
Claim & Measured \\
\midrule
Plain bar lowered in every flooded cell & True \\
Plain bar delta over flood (median) & -0.178 \\
Floor attenuates in every flooded cell & True \\
Floored bar delta over flood (median) & -0.054 \\
HWM bar non-decreasing in every cell & True \\
Flood earns wealth in any cell & False \\
Streams with a post-flood certification (plain / floored / hwm) & 4 / 1 / 1 \\
Flood buys additional runway in any cell & False \\
Cells stopped before injection (evaporation) & 36 of 36 \\
\bottomrule
\end{tabular}

\end{table}

\FloatBarrier
\subsection{Effective ledger size in these searches}
\label{sec:results-neff}

\begin{figure}[htbp]
\centering
\includegraphics[width=\columnwidth]{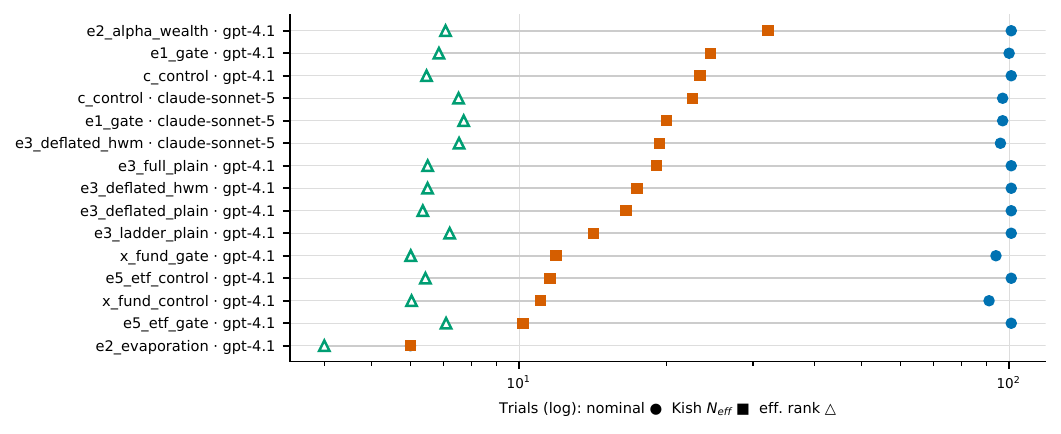}
\caption{Nominal trial count (circles) against Kish effective size (squares)
and eigenvalue effective rank (triangles), median per cell, on a logarithmic
axis. In these searches, a hundred trials correspond to 6 to 8 effectively
independent return streams.}
\Description{Dot plot per experimental cell on a logarithmic axis. Nominal
trial counts cluster near one hundred, Kish effective sizes range from 10 to
32, and eigenvalue effective ranks range from 6 to 8. The early-stopped
evaporation cell sits below these ranges.}
\label{fig:neff}
\end{figure}

Figure~\ref{fig:neff} reports the dependence diagnostic of Section~\ref{sec:anatomy-neff}, computed identically for every arm from each run's rebuilt candidate-return matrix. The result is uniform across cells, models, and universes: a nominal ledger of about 100 trials corresponds to a Kish effective size of 10 to 32 and an eigenvalue effective rank of 6 to 8. For these search trajectories, in other words, a hundred candidates amount to only a handful of effectively independent return streams, and to 22 to 44 distinct hypotheses by the key-based endpoint of Table~\ref{tab:endpoints}. The corresponding gap between the nominal and effective bars is 0.19 to 0.36 in annualised Sharpe. We read this as a statement about these searches rather than about the finance literature at large: within these trajectories, the nominal trial count substantially overstates the diversity of the hypotheses tried, so a bar computed at the nominal count is correspondingly conservative. Because the live bar deliberately uses the nominal count (the reasoning of Section~\ref{sec:anatomy-neff}), this conservatism is a measured and intended property rather than a hidden error, and a less conservative but still robust in-loop treatment is better sought through dependence-aware e-value methods than through a gameable point estimate.

\FloatBarrier
\subsection{Robustness to the frozen hyperparameters}
\label{sec:results-sensitivity}

The comparisons above rest on a small number of seeds, so we ask separately whether the qualitative findings depend on the specific frozen hyperparameters. Because the recorded candidate trajectories are fixed, we can vary the variance floor, the stopping parameters, the duplicate threshold, and the injected flood size, and re-derive each finding over the same searches without new runs. These are counterfactuals over the recorded trajectories rather than new experiments, and they cannot vary the ladder step, which would change what the critic sees and hence the proposals; that case is left to future work. Table~\ref{tab:sensitivity} collects the results, which we read as directional support rather than as significance tests.

The two central findings are insensitive to their parameters. The gate's honest-empty count is 36 of 37 at every variance floor from the fifth to the fiftieth percentile of the calibration pool, a floor range of roughly $0.12$ to $0.19$, and the single certification persists throughout; the floor binds on only 5 of the 37 gate runs at the pre-registered tenth percentile, because most searches already exceed it in trial-Sharpe dispersion. The adaptive budget reaches the iteration cap on all 34 full-length trajectories for every reward $\psi \in \{10, 20, 40\}$, so the finding that budget adaptivity adds nothing over a fixed budget does not depend on the reward size. Evaporation stopping halts far below the 100-candidate budget for every patience $m \in \{2, 3, 5\}$ (median stop after 1, 2, and 7 refinement candidates, with maxima of 5, 16, and 18), and larger $m$ halts later as expected, so the early-stopping finding is not an artifact of $m = 3$, though the exact trial count varies across trajectories. In the dilution stress test, a larger flood lowers the plain bar further, by $0.11$, $0.19$, and $0.24$ in annualised Sharpe for floods of 15, 30, and 45 near-mean trials, while the monotone bar does not move and the flood earns no wealth at any size; this ablation replays one representative stream, so its 30-trial value differs slightly from the twelve-stream median of Table~\ref{tab:e4}.

The feedback effect of Section~\ref{sec:results-e3} is the most parameter-dependent, which is consistent with its being our weakest result. The excess near-duplication of deflated over full feedback under the plain bar is non-negative at all three correlation thresholds ($+0.00$, $+0.04$, and $+0.01$ at $\rho \in \{0.90, 0.95, 0.99\}$), consistent with the predicted direction but with a magnitude that is clearest at the pre-registered $0.95$; under the monotone bar the excess disappears at every threshold, with the deflated rate within $0.01$ of the full-feedback baseline at $\rho \in \{0.90, 0.95\}$ and $0.03$ below it at $0.99$, consistent with the removal of the downward gradient. We therefore treat the feedback result as a small and directionally consistent effect, and rest the paper's conclusions on the two findings that hold across their whole grids.

\begin{table}[htbp]
\centering
\caption{Sensitivity of the qualitative findings to the frozen hyperparameters
(exploratory, deterministic, no new runs). Each row varies one parameter and
re-derives the finding over the recorded trajectories.}
\label{tab:sensitivity}
\footnotesize
\begin{tabular}{ll}
\toprule
Ablation & Result across the grid \\
\midrule
Variance floor (gate honest-empty / 37) & p5: 36; p10: 36; p25: 36; p50: 36 \\
Evaporation stop (median refinement candidates) & m=2: 1; m=3: 2; m=5: 7 \\
Alpha-wealth stop (median refinement candidates) & $\psi$=10: 100; $\psi$=20: 100; $\psi$=40: 100 \\
E3 deflated-minus-full dup rate (plain) & $\rho$=.90: +0.00; $\rho$=.95: +0.04; $\rho$=.99: +0.01 \\
Flood: plain-bar delta / hwm monotone & n=15: -0.11/True; n=30: -0.19/True; n=45: -0.24/True \\
\bottomrule
\end{tabular}

\end{table}

\FloatBarrier
\section{Discussion}
\label{sec:discussion}

\paragraph{What in-loop correction changed here.}
Moving the correction inside the loop did not uncover more edge; on the markets and features we searched, under these instruments, there was no active edge to uncover, which is consistent with the post-hoc verdict of our prior system. What it changed lies elsewhere. First, the system's own report, rather than a later audit, states that nothing certifies, which turns a common failure mode of the discovery literature, reporting an inflated winner, into an outcome the system tends not to produce. Second, on these runs the honest verdict was also the inexpensive one, because the same statistic that declines to certify also signals early that continuing is not worthwhile. A loop that cannot afford to reject its hundredth candidate after the fact can afford a bar that stops it at its eighth. We state both effects as properties of the analysed system.

\paragraph{A design lesson that may generalise.}
The one point we would carry beyond trading is methodological: a statistic placed inside a search loop becomes something the search can optimise against, and its incentive structure is worth analysing before deployment rather than after. Our dilution analysis found the incentive on paper, the feedback study observed a frontier model shifting in the predicted direction, by a small margin, when shown the gradient, and the benchmark-audit literature reports the analogous dynamic when the statistic is a test suite \citep{Zhong2025impossiblebench,Wang2026benchjack}. Where such a loop is unavoidable, two of our choices may transfer: prefer instruments that cannot be pushed in the exploitable direction, as the monotone bar cannot be pushed down; and keep dependence corrections as after-the-fact diagnostics rather than live dials, since a correction that responds to correlation can be driven by manufactured correlation. We offer these as design suggestions supported by our experiments, not as theorems.

\paragraph{On effective size.}
That a hundred candidates amount to 6 to 8 effectively independent return streams, in these searches, has two readings, and we hold both loosely. It suggests that a bar computed at the nominal count is conservative for such trajectories, and it suggests that the searches are less exploratory than their trial counts imply, proposing variations more than distinct hypotheses. Both readings are relevant to the alpha-mining systems that report ever larger candidate counts \citep{Tang2025alphaagent,Li2025rdagentq,Han2026quantaalpha}, in that the count that inflates a backtest and the diversity that would justify a search budget are different quantities and neither equals the number of candidates. A dependence-aware e-value construction \citep{RamdasWang2025,Zhang2025egai} is a natural path to an in-loop correction that is neither as conservative as the nominal count nor as gameable as a live effective count, and our measurements indicate the constraints it would need to satisfy.

\paragraph{Suggested defaults.}
For builders of autonomous discovery systems, the configuration our study supports as a starting point is concrete: a deflated certification gate on a floored, monotone bar; state-based stopping of the evaporation kind; quantised or hardened-bar feedback to the proposer; the nominal count in the loop with the effective count reported beside each result; and bit-replayable accounting, so that the claim that a search was honest can be checked rather than asserted. We present these as defaults that worked here, to be revisited as the setting changes.

\section{Limitations}
\label{sec:limitations}

Several limitations bound the claims. The comparisons rest on three seeds per cell, and five on the headline pair, so we report distributions and directions rather than significance, and the feedback sign pattern in particular is a consistent direction across seeds rather than a tested effect. To partly offset the limited seed count we re-derived each finding over the recorded trajectories across a grid of the frozen hyperparameters (Section~\ref{sec:results-sensitivity}); the two central findings are invariant across that grid, while the feedback effect is small and threshold-dependent, and we weight the conclusions accordingly. Default-temperature sampling also produces convergent searches, with the Claude model repeatedly selecting identical strategies across seeds, which narrows seed dispersion in ways we report rather than smooth over. The scope is a single harness lineage, two frontier models, and daily equity and ETF data with one realistic cost model inherited from the prior system; the exploratory fundamentals cell uses annual-cadence filings and is labelled accordingly. Because the setting is single-market and daily-frequency, the finding that nothing active certifies is a statement about this search space, not about markets in general. On guarantees, the alpha-wealth rule carries no false-discovery-rate guarantee under the dependence of strategy search, and we claim none; the bar controls an estimated expected maximum under an approximate-normality assumption, not a formal error rate for the composite procedure, and Proposition~\ref{prop:dilution} is a first-order statement under stated approximations. Treatment arms share the control's data and seed plans by design, to isolate single-axis effects, so the arms are not independent replications of one another. Finally, the finding that dilution does not fund itself is specific to the frozen budget and to these trial streams; the mechanical channel exists, and other budget settings could change the balance. With 53 marginal 95\% intervals, a few spurious zero-exclusions would be unremarkable under a global null; none occurred. The intervals' median half-width is about $0.9$ in annualised Sharpe, so their inclusion of zero rules out only large edges on this window, and we read it as absence of contradiction rather than evidence of absence. Finally, the suite contains no planted true edge, so the false-negative cost of the gate and its defences is not measured here; the single certification is the only observed pass.

\section{Conclusion}
\label{sec:conclusion}

We placed a multiplicity correction inside an autonomous search loop and studied the consequences on both sides of the loop. On the searcher's side, a live bar is an object the search can act upon: a near-mean trial lowers it, a budgeted stopping rule could in principle help fund the exploit, and a transparent critic can move in the predicted direction, so in-loop honesty is partly a matter of designing the metric against being gamed, which a variance floor and a monotone bar address at a measured cost. On the reporting side, across 53 audited searches on two models, two universes, and two feature spaces, the gated system almost never vouched for what its own search produced, reporting 36 honest empties out of 37 with one discriminating certification, while controls vouched every time, and no out-of-sample interval contradicted the gate's reticence. A simple state-based stopping rule reached the same outcome for about a twelfth of the trial budget on these runs, and a dependence diagnostic indicated that these searches explore far fewer effectively independent return streams than their nominal trial counts suggest. Moved inside the loop, the instruments did not manufacture discoveries; they made the search report honestly, and stop efficiently, when there was nothing to find. For autonomous discovery systems whose evaluations are part of their own environment, we offer the analysis, architecture, and protocol here as one worked template, developed in a single trading system, for building loops whose honesty can be checked rather than promised.

\paragraph{Reproducibility.}
A public archive released with the paper contains the in-loop accounting layer under study (the objective, stopping, and feedback registries with the ledger state they operate on), the statistical instruments, the replay auditor that recomputes every in-loop decision of a run from its recorded trace, the frozen configurations, calibration and audit records, endpoint extraction, the deterministic figure and table generator, and the complete recorded traces of every reported run. Every number in Tables~\ref{tab:endpoints} and \ref{tab:e4} and Figures~\ref{fig:forest} to \ref{fig:neff} regenerates from these sources, and every in-loop decision of every reported run replays exactly from its trace. The archive is published on Zenodo at \url{https://doi.org/10.5281/zenodo.21725589}, a concept DOI that resolves to its latest version. The full discovery platform is available from the author on reasonable request for research verification purposes.

\begin{acks}
This research received no specific grant from any funding agency in the
public, commercial, or not-for-profit sectors.
\end{acks}

\bibliographystyle{ACM-Reference-Format}
\bibliography{references}

\end{document}